# X-ray magnetic circular dichroism study of epitaxial magnetite ultrathin film on MgO (100)

W. Q. Liu,[1,2] M. Y. Song,[3] N. J. Maltby,[2] S. P. Li,[2] J. G. Lin,[3] M. G. Samant,[4] S. S. P. Parkin,[4] P. Bencok,[5] Paul Steadman,[5] Alexey Dobrynin,[5] Y. B. Xu,[1,2,*] R. Zhang[1,*]

[1] York-Nanjing International Center for Spintronics (YNICS), School of Electronics Science and Engineering, Nanjing University, Nanjing, 210093, China

[2] Spintronics and Nanodevice Laboratory, Department of Electronics, University of York, York YO10 5DD, UK

[3] Center for Condensed Matter Sciences, National Taiwan University, Taipei 106, Taiwan

[4] IBM Research Division, Almaden Research Center, San Jose, CA 95120 USA

[5] Diamond Light Source, Didcot OX11 0DE, UK

**Abstract:** The spin and orbital magnetic moments of the $Fe_3O_4$ epitaxial ultrathin film synthesized by plasma assisted simultaneous oxidization on MgO(100) have been studied with X-ray magnetic circular dichroism (XMCD). The ultrathin film retains a rather large total magnetic moment, i.e. $(2.73\pm0.15)\mu_B$/f.u., which is ~ 70% of that for the bulk-like $Fe_3O_4$. A significant unquenched orbital moment up to $(0.54\pm0.05)$ $\mu_B$/f.u. was observed, which could come from the symmetry breaking at the $Fe_3O_4$/MgO interface. Such sizable orbital moment will add capacities to the $Fe_3O_4$-based spintronics devices in the magnetization reversal by the electric field.

*I. Background*

Magnetite, or $Fe_3O_4$, is a competitive contender in the race to become the one of the key materials in the future spintronics computing, or the spin-operation-based data processing and sensing, due to its high spin polarization near the Fermi level ($E_F$). The experimentally true half-metallic state was reported by Dedkov *et al.*[1] by means of spin and angle-resolved photoemission spectroscopy, from which P = - (80 ± 5)% was obtained at $E_F$, consistent with the spin-split band energies from DFT calculations.[2] More desirably, the high Curie temperature ($T_c$) of $Fe_3O_4$ makes it a promising candidate for room temperature use. Fascinating properties of spin transport have also been presented in $Fe_3O_4$, i.e. spin Seebeck effect,[3] spin filter effect,[4] gate voltage-induced phase transition,[5] and spin valve effect of $Fe_3O_4$/MgO/$Fe_3O_4$ junctions.[6,7] Yet at the meantime, many fundamental properties of magnetite such as the half-metallicity, spin and orbital ordering, Verwey transition mechanism and the coupling mechanism between different sites have long been open issues, and with the thickness down to nanometer scale, these issues become even more sophisticated.

The rather complicated magnetic structure of $Fe_3O_4$ was partly proposed by Verwey and Hayman in 1941[8] and the total structure was put forward by Neel in 1948,[9] and then confirmed three years later by neutron scattering.[10] $Fe_3O_4$ has cubic inverse spinel structure, where $Fe^{3+}$ ions occupy tetrahedral sites (usually called A sites), whereas octahedral sites (B sites) are occupied by both $Fe^{3+}$ and $Fe^{2+}$ ions. The spin of $Fe^{3+}$ ions at octahedral and tetrahedral sites are aligned antiparallel to each other leading to a net spin magnetic moment ($m_{spin}$) of 4 $\mu_B$/f.u., corresponding to a fully occupied local majority band (opposite for A and B sites). The presence of integer $m_{spin}$ and vanishing orbital moment ($m_{orb}$) of magnetite are expected as an indication for a B-site minority electron conduction mechanism, and its accompanied full spin polarization at the $E_F$. However, concerning with that, controversial results have been reported utilizing techniques such superconducting quantum interference device (SQUID) magnetometer,[11,38] XMCD,[12,27,30,39,38] and magnetic Compton scattering (MCS)[33,34] and calculations with local density approximation (LDA),[30] LDA+U,[30] local spin density approximation (LSDA)+U,[2] and moment analysis etc..[12] The fundamental magnetic properties of $Fe_3O_4$ show strong dependence on the sample preparation methodology. By simultaneous oxidation, Babu *et al.*[32] observed $m_{spin}$ = (1.20±0.05) $\mu_B$/f.u. of $Fe_3O_4$ on $BaTiO_3$, and enhanced $m_{spin}$ = 7.7 $\mu_B$/f.u. was reported by Arora *et al.*.[11] Whilst using pulsed laser deposition (PLD), Orna *et al.*[38] observed greatly reduced $m_{spin}$ = 1.83 $\mu_B$/f.u. of $Fe_3O_4$ thin film on MgO. And using floating zone methods, the reported $m_{orb}$ of bulk $Fe_3O_4$ varies from (0.67±0.09) $\mu_B$/f.u. by Huang et al.,[30] (0.51±0.05) $\mu_B$/f.u. by Li et al.,[33] to (0.06±0.14) $\mu_B$/f.u. by Duffy et al.[34] and all the way down to –0.001$\mu_B$/f.u. by Goering et al..[27] Theoretical analyses were presented with even sharper contrast, varying from 0.43 $\mu_B$/f.u. by Huang et al.[30] to 0.02 $\mu_B$/f.u. by Antonov et al..[2] The well-known Verwey transition of magnetite is accompanied by a transition to a low symmetry structure, by which the $m_{spin}$ and especially $m_{orb}$ are expected to change significantly. The experimental work,[11,34,38] however, has so far found no difference of them across the transition, which questions the picture of a fully A site $Fe^{3+}$ and a mixed-valence B site configuration of magnetite.

$Fe_3O_4$/MgO heterostructure has attracted great interest within the spintronics community in the recent few years. Crystalline MgO-based magnetic tunnel junctions have achieved remarkable success in assisting efficient spin injections for various applications.[13,14,15] An insulating layer of MgO can be used as a tunneling barrier, which not only relieves the conductivity mismatch problem but also works as a spin filter. Moreover, MgO forms an excellent diffusion barrier with thermal stability up to 800 °C, effectively preventing the intermixing at a given ferromagnet-semiconductor interface. Efforts have been made to explore the magnetic behavior of nanoscale epitaxial thin films on bulk MgO(100).[11,16] Yet to fundamentally understand character of $3d$ electrons in $Fe_3O_4$, it is a prerequisite to determine the $m_{spin}$ and $m_{orb}$, respectively and unambiguously. Recently, we have reported a unquenched $m_{orb}$ and a well retained $m_{spin}$ in $Fe_3O_4$ thin films grown by post-growth annealing method.[17] In this paper, we furthermore present a XMCD study of $Fe_3O_4$ in a distinct configuration, prepared by plasma-assisted simultaneous oxidation, aiming to have an insight of the different magnetic properties of the $Fe_3O_4$ thin film caused by the substrate and preparation methodology, and to contribute to the open question of the magnetic moments of the $Fe_3O_4$ in nano scale regime.

## *II. Sample preparation and global characterization*

The magnetite ultrathin film used in this study was grown by oxygen-plasma-assisted MBE at a substrate temperature of 623 K and in oxygen pressure of 2.5 × $10^{-7}$ torr. Prior to the growth, the MgO was annealed at 600 °C and after 30 minutes, a sharp MgO(100) was obtained as can be seen from the reflection high-energy electron diffraction (RHEED) pattern in Fig. 1. (a). The 12 nm $Fe_3O_4$ was then deposited by effusion cells at a rate of 0.04 Å per second at 350 °C. The comparatively large oxygen ions form a face-centered-cubic (fcc) lattice and the Fe atoms are located in interstitial sites as shown in Fig. 1. (b).

The resistance versus temperature (R-T) relationship of the 12 nm $Fe_3O_4$/MgO thin film was measured by patterning into a standard Hall bar geometry. As shown in Fig. 2, a discontinuous change of the resistivity was observed at ~ 97 K, corresponding to the Verwey transition. Ferromagnetic resonance (FMR) measurement was performed and published elsewhere, by which different resonance features was observed below and above the transition.[16]

The macroscopic magnetization of the sample was characterized by means of SQUID magnetometer. Figure 3. (a) presents the magnetization versus temperature relationship (M-T), obtained by cooling the sample from 300 to 20 K in zero magnetic field, followed by an application of a static magnetic field of 100 Oe and recording the magnetization values during the warming cycle to 300 K. The magnetization drops with the decreasing temperature and the Verwey transition at ~ 97 K, consistent with the transport measurement, can be distinguished from the dM/dT (see the Fig. 3(a) inset). The magnetization versus field (M-H) loops in Fig. 3. (b) was obtained by applying the magnetic field in the film plane and with the subtraction of a small diamagnetic contribution from the sample holder. It can be seen that the sample saturates at ~ 500 Oe with a magnetization of ~ 270 emu/cm$^3$, which is ~ 70% of the principle bulk-like magnetite saturation value, i.e., 480 emu/cm$^3$ or 4. 0 $\mu_B$/f.u..

It is generally accepted that the presence of Verwey transition is very sensitive to the stoichiometry and homogeneity of magnetite thin films.[18] The temperature ($T_V$) at which such a transition occurs has been commonly observed to decrease from the bulk value down to 85K and could even disappear with the decreasing $Fe_3O_4$ thicknesse.[11, 19, 20, 21, 38] The observed transition temperature of the 12nm $Fe_3O_4$/MgO(100) thin film in this study agrees well with the reported $T_V$ of the magnetite films on or near stoichiometry,[11, 22] suggesting the sample is not overly suffering from cation or anion vacancies.

## III. XMCD measurement

X-ray absorption spectroscopy (XAS) and XMCD experiments at the Fe $L_{2,3}$ absorption edges were performed at the beamline I10 of Diamond Light Source, UK. Circularly polarized X-rays with 100% degree of polarization[23] were used in normal incidence with respect to the sample plane and parallel with the applied magnetic field, in order to minimize the nonmagnetic asymmetries, as shown in Fig. 4. The XAS spectra were obtained by total electron yield (TEY) detection. The XMCD was taken as the difference of the XAS spectra, i.e., $\sigma^- - \sigma^+$, obtained by flipping the X-ray helicity at a fixed magnetic field of 30 kOe, under which the sample is fully magnetized with little paramagnetic contribution. Typical XAS and XMCD spectra at 300 K of the 1 2nm $Fe_3O_4$/MgO(100) are presented in Fig. 5. The XMCD spectra well reproduce the features of those from the theoretical calculations and previous observations of the ferrimagnetic $Fe_3O_4$ with the three Fe alternative up and down peaks coming from the antiparallel spin orientations of the A and B sites. The relative intensities of these $L_3$ peaks show lesser occupation of $Fe^{2+}$ B sites, which may be associated with the slight formation of a secondary $\gamma$-$Fe_2O_3$ phase.[24,25] The complex form of the XMCD spectrum arises because of an overlap of different sets of multiplet structures. The B sites $Fe^{3+}$ and $Fe^{2+}$ spin-up states exhibit negative peaks at Fe $L_3$ edge and positive peaks at the Fe $L_2$ edge, while the A sites $Fe^{3+}$ spin-down states behave the oppositely at the Fe $L_3$ and $L_2$ edges, respectively.

The $m_{spin}$ and $m_{orb}$ of the the 12 nm $Fe_3O_4$/MgO(100) were calculated by applying sum rules on the integrated XMCD and total XAS spectra of Fe $L_{2,3}$ edges based on equation (1),[26]

$$<L_z> = -\frac{4}{3}n_h \frac{\int_{L_{2,3}} (\sigma^+ - \sigma^-)\, dE}{\int_{L_{2,3}} (\sigma^+ + \sigma^-)\, dE}$$

$$<S_z> = -n_h \frac{6\int_{L_3} (\sigma^+ - \sigma^-)\, dE - 4\int_{L_{2,3}} (\sigma^+ - \sigma^-)\, dE}{\int_{L_{2,3}} (\sigma^+ + \sigma^-)\, dE} \times SC + <T_z> \quad (1)$$

where $n_h$ - the effective number of $3d$-band holes had been taken from literature.[30] In order to rule out non-magnetic parts of the XAS spectra an arctangent based step function is used to fit the threshold.[26] As can be seen from Fig. 2, unlike an infinite 'long tail' reported by Goering *et al.*,[27] the integrated spectrum of both XMCD and total XAS of our data quickly saturate at ~ 726 eV. Therefore an integration range to 730 eV is sufficient, giving $m_{spin}$ = (2.19±0.1) $\mu_B$/f.u., $m_{orb}$ = (0.54±0.05) $\mu_B$/f.u. and the $m_{orb}$ / $m_{spin}$ ratio as large as 0.25. It should be noted that all sum rules-related values given here are the average information over the whole formula unit (f.u.) of the three cations contribution. Possible artefact of the experimental set up and data analysis of XMCD of magnetite were discussed in detail by Goering *et al.*.[27] In general, the nonmagenetic part of the raw data is smaller than 1/1000 of the total absorption. The saturation effect in our case is estimated to be about 3% in the normal incidence configuration. Besides, the magnetic dipole term $<T_z>$ plays a rather insignificant role because of the predominantly cubic symmetry of magnetite, even under a scenario of additional surface symmetry breaking. Besides, the good agreement of the $m_{s+l}$ obtained from SQUID measurement with that calculated from XMCD is an additional proof of the proper application of the sum rules in this study.

*IV. Discussion*

Unquenched $m_{orb}$, or strong spin-orbit coupling $(<LS>)$, is a desired property in terms of the controllability by electric field in spintronics operation[28], however, which have been reported with controversy in magnetite. Early theoretical work based on the picture of bulk $Fe_3O_4$ possessing $m_{spin}$ = 4.0 $\mu_B$/f.u. and nearly vanishing $m_{orb}$. McQueeney *et al.*[29] obtained a high $<LS>$ of magnetite of the order of 10 meV, pointing to a large $m_{orb}$ to expect. The XMCD performed by Huang *et al.*[30] suggests a large unquenched $m_{orb}$, typically 0.67 $\mu_B$/f.u. along with a $m_{spin}$ = 3.68 $\mu_B$/f.u. at the temperature both above and below Verwey transition. These results are well resembled by calculations using the LDA + U scheme.[30] The large $m_{orb}$ has been attributed to a strong on-site Coulomb interaction and corresponding 3*d* correlation effects. Similarly sizable $m_{orb}$ was also observed by Kang *et al.*[31] in Mn substitution at the A site, which changes the valence of the B-site Fe and by Babu *et al.*[32] in ultrathin $Fe_3O_4$ on $BaTiO_3$(001). By sharply contrast, XMCD performed by Goering *et al.*[27] suggests that there is in fact a vanishingly small $m_{orb}$ on the Fe sites. To avoid the systematic errors arises from the XMCD data analysis, MCS were performed, which still end in controversial results. Non-integral $m_{spin}$ = 3.54 $\mu_B$/f.u. and correspondingly $m_{orb}$ = 0.51 $\mu_B$/f.u. were observed by Li *et al.*,[33] while Duffy *et al.*[34] reported again nearly vanishing $m_{orb}$. Goering *et al.*[12] has recently tried to explain the large variety of published results by the independent analysis of the Fe $L_{2,3}$ edge XAS, moment analysis fit of the Fe $L_{2,3}$ edge XMCD, and by the comparison with O K edge XMCD. In consistent with Goering's, our data also exhibit an intensity ratio $r_{23}$ = 0.25, strongly reduced from a pure statistical case where $r_{23}$ = 0.5, which support the presence of large $m_{orb}$ and respective $<LS>$ expectation values. According to Goering's argument, these orbital moments are located at the A and B sites of magnetite and aligned antiparallel with each other, similar to the spin moments, though quantitatively, this scenario questions the picture of a fully A site $Fe^{3+}$ and a mixed-valent B site configuration. Table 1 summaries some of the experimental and theoretical efforts within the thin film regime toward this issue. Regardless the controversial reports on various form of magnetite, our results (the first line of the table) suggest the existence of a large unquenched $m_{orb}$ with $m_{orb}$ / $m_{spin}$ = 0.25 in the epitaxial $Fe_3O_4$ ultrathin film on MgO(100). While the unquenched $m_{orb}$ might be a intrinsic property of the bulk-like $Fe_3O_4$, which is still a hotly debated topic, our result could also originates from modification of crystal lattice symmetry as $m_{orb}$ in the low dimensional magnetic systems can be strongly enhanced by the reduced symmetry of the crystal field as found in the Fe/GaAs(100) system.[35]

It is worth noting that not only the nanoscale full epitaxial $Fe_3O_4$/MgO(100) heterostructure exhibits considerably large $m_{orb}$, its total moment ($m_{total}$) also retains ~ 70% of the bulk value. The deviation of the magnetic moment of thin films from the bulk-like $Fe_3O_4$ is usually attributed to three forms of missing compensation or symmetry breaking. The first one is the formation of antiphase boundaries (APBs) raised from the epitaxy growth process due to the fact that $Fe_3O_4$ has twice the unit-cell size of MgO.[36,37] In magnetite thin films, the magnetic interactions are altered at the APBs, across which the intrasublattice exchange interactions dominate, reversing the spin coupling. Therefore, the structural boundary separates oppositely magnetized regions and the resultant coupling between two domains turns out to be either frustrated or antiferromagnetic. Such antiferromagnetic exchange interactions usually lead to saturation fields as large as 70 kOe,[36] but can be neglected for our sample, whose magnetization saturate before 500 Oe (see Fig. 3(b)). The second mechanism of non-compensation occurs due to the less cubic symmetry of magnetite at the surface and the interface of $Fe_3O_4$/MgO. Among the very few work performed on $Fe_3O_4$ thin films, Orna et al.[38] reported significantly shrinking $m_{spin}$ = 1.83 $\mu_B$/f.u. in $Fe_3O_4$(8 nm)/MgO, as well as the observation by Babu et al.[32] of $m_{spin}$ = (1.20±0.05) $\mu_B$/f.u. in $Fe_3O_4$ (2.5 nm)/$BaTiO_3$. Even in the bulk, strongly reduced $m_{spin}$ down to (1.73±0.02) $\mu_B$/f.u. was observed by Goering et al..[39] By contrast again, large $m_{spin}$ of 7.7 $\mu_B$/f.u in $Fe_3O_4$(5 nm)/MgO was reported by Arora et al.,[11] who attributed the enhancement to the uncompensated spin between A and B sublattices at the surface and across the APBs. However, this enhancement may also come from the magnetic impurities as suggested by Orna et al..[38] The inter-diffusion of ions, which tends to substitute onto B-sites, is the third possibility. Although our sample were grown at a moderate growth temperature (350 °C), one may still predict an appreciable inter-diffusion given the *ex situ* measurement in the study were not carried out immediately after the growth. However, if any, such substitution would only happen at the first 1-2 atom layers at the $Fe_3O_4$/MgO interface. Therefore results presented in this paper are rather representative for a $Fe_3O_4$ thin film on MgO without the mixture of $Mg^{2+}$ ions.

*IV. Conclusion*

To summarize, we have performed XMCD of a $Fe_3O_4$ epitaxial thin film on MgO(100) synthesized by plasma-assisted simultaneous oxidization MBE process. High quality XAS and XMCD spectra were obtained and carefully analysed with the sum rules. A significant unquenched $m_{orb}$ was observed, which may be an intrinsic property of the $Fe_3O_4$ or come from the symmetry breaking at the $Fe_3O_4$/MgO(100) interface. Such sizable $m_{orb}$ has strong implications for adding capacities to spintronics devices, since high *<LS>* coupling is mandatory for realizing the ultrafast switching of spin polarization by electric field and circularly polarized light. Moreover, our 12 nm $Fe_3O_4$/MgO(100) heterostructure retains a large $m_{total}$ of ~ 70% of the bulk-like stoichiometric $Fe_3O_4$. Based on the presence of Verwey transition, such large magnetic moments can stay independent from the cation or anion vacancies and the APBs, if any. Our results offer direct experimental evidence for addressing the open issue of the spin and orbital moments of magnetite, particularly, in its epitaxial ultrathin film form, which is significant for achieving high efficient spin injection and electrical spin manipulation in the full epitaxial spintronic heterostructures.

**Acknowledgements**

This work is supported by the State Key Programme for Basic Research of China (Grants No. 2014CB921101), NSFC (Grants No. 61274102) and PAPD project, UK EPSRC and STFC. We acknowledge Jill S. Weaver for sharing the computation code of the sum rule calculation.

**Tables**

| Sample | Method | $m_{orb}$ ($\mu_B$/f.u.) | $m_{spin}$ ($\mu_B$/f.u.) | $m_{total}$ ($\mu_B$/f.u.) | $m_{orb}/m_{spin}$ | *Ref.* |
|---|---|---|---|---|---|---|
| 12 nm $Fe_3O_4$/MgO(100) | XMCD | 0.54±0.05 | 2.19±0.1 | 2.73±0.15 | 0.25 | * |
| 8 nm $Fe_3O_4$/MgO/GaAs(100) | XMCD | 0.47±0.05 | 2.84±0.1 | 3.32±0.15 | 0.17 | 17 |
| 5 nm $Fe_3O_4$/MgO(001) | SQUID | | | 7.7 | | 11 |
| 8 nm $Fe_3O_4$/MgO(001) | XMCD | | | 1.83 | <0.05 | 38 |
| 2.5 nm $Fe_3O_4$/$BaTiO_3$(001) | XMCD | 0.44±0.05 | 1.20±0.05 | 1.64 | 0.37 | 32 |
| Theory | LDA | 0.06 | 4.0 | 4.06 | 0.015 | 30 |
| Theory | LDA+U | 0.43 | 4.0 | 4.43 | 0.108 | 30 |
| Theory | LSDA+U | 0.02 | 3.7 | 3.72 | 0.005 | 2 |

**Table 1.** The experimental and theoretical reports of the magnetic moments of magnetite thin films and the results of the present study (marked as *).

**Figures**

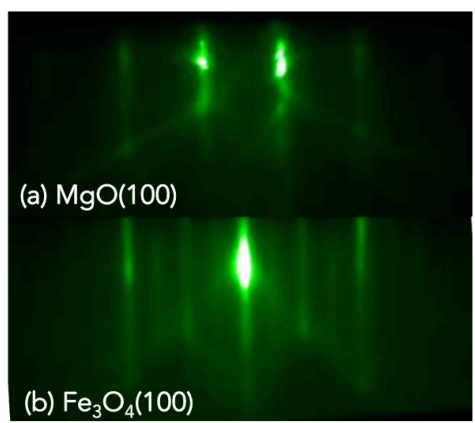

**Figure 1.** Typical RHEED patterns of (a) MgO(100) after annealing, (b) $Fe_3O_4$/MgO(100), taking during the MBE growth.

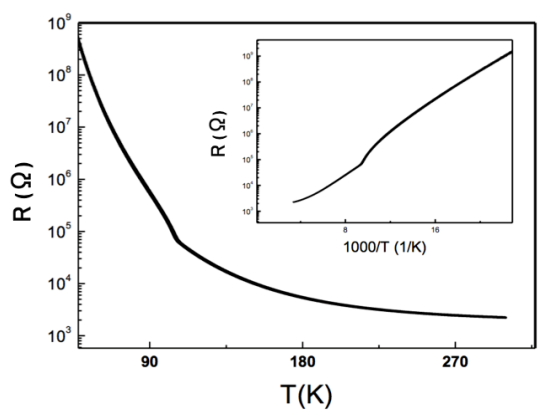

**Figure 2.** The resistance versus temperature relationship. Inset: the resistivity versus the inverse of temperature.

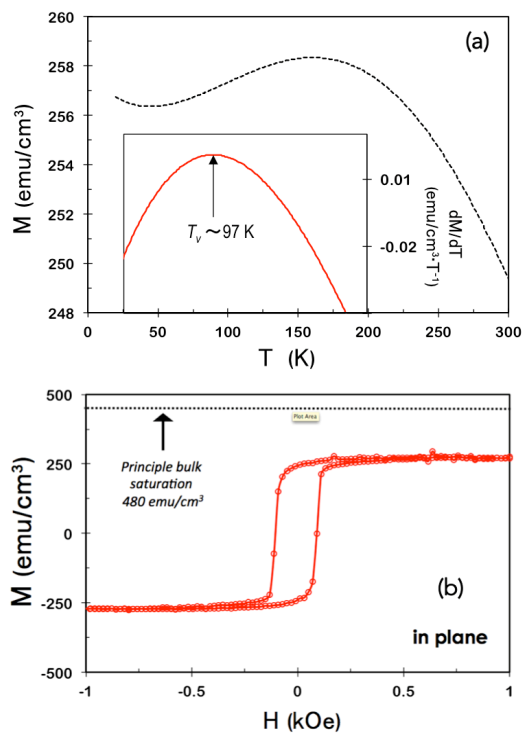

**Figure 3.** The global magnetization versus (a) temperature. Inset: the gradient of magnetization versus temperature. (b) applied magnetic field. Dash line indicates the value of bulk saturation.

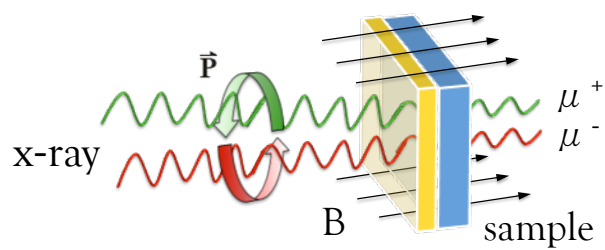

**Figure 4.** A schematic diagram of the XAS and XMCD experimental configuration utilized in this study.

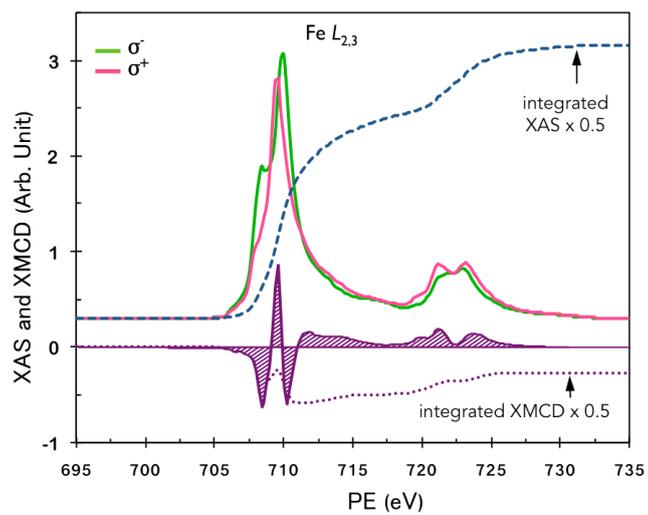

**Figure 5.** A typical XAS pair and XMCD spectra of the magnetite thin film sample obtained at 30 kOe and room temperature.